\documentclass[aps,prd,preprint]{revtex4}
\usepackage{amssymb}
\usepackage{amsfonts}
\usepackage{amsmath}
\usepackage{graphicx}

\begin{document}

\title{Hawking Radiation of Spin-1 Particles From Three Dimensional Rotating
Hairy Black Hole }
\author{I. Sakalli$^{\ast}$}
\author{A. Ovgun$^{\ddag}$}
\affiliation{Department of Physics, Eastern Mediterranean University,}
\affiliation{G. Magusa, North Cyprus, Mersin-10, Turkey}
\affiliation{$^{\ast}$izzet.sakalli@emu.edu.tr}
\affiliation{$^{\dagger}$ali.ovgun@emu.edu.tr}

\date{\today}
\begin{abstract}
In the present article, we study the Hawking radiation (HR) of spin$-1$
particles -- so-called vector particles -- from a three dimensional ($3D$)\
rotating black hole with scalar hair (RBHWSH) using Hamilton-Jacobi (HJ)
ansatz. Putting the Proca equation amalgamated with the WKB\ approximation
in process, the tunneling spectrum of vector particles is obtained. We
recover the standard Hawking temperature corresponding to the emission of
these particles from RBHWSH.
\end{abstract}

\maketitle

\section{Introduction}

One of the most radical predictions of general relativity (GR) is the
existence of black holes (BHs). According to the seminal works of Hawking 
\cite{Hawking1,Hawking2,Hawking3}, BHs are not entirely black. That was the
surprising claim made by Hawking over forty years ago. Examining the
behavior of quantum fluctuations around the event horizon of a BH, Hawking
substantiated the theory that BHs radiate thermal radiation, with a constant
temperature -- so-called Hawking temperature -- directly proportional to the
surface gravity $\kappa$, which is the gravitational acceleration
experienced at the BH's horizon:

\begin{equation}
T_{H}=\frac{\hslash\kappa}{2\pi},   \label{1}
\end{equation}

with the system of units $c=G=k_{B}=1$. The works of Hawking and Bekenstein 
\cite{Bekenstein} -- and of others \cite%
{Ruffini,Unruh,Brout,Kraus,Corley,Corley2,Visser,Srinivasan,Parikh,Visser2,Kiefer}%
, rederiving $T_{H}$ in various ways -- brings together the normally
disparate areas of GR, quantum mechanics (QM), and thermodynamics.
Enthusiasm for understanding of the underlying coordinations between these
subjects of physics creates ample motivation for the study of HR (see for
instance \cite%
{Wald,Page,Singleton,Majhi,Zhang,Sakalli1,Vanzo,Sakalli2,Page2,Sakalli3,Modak,Hawking4}
and references therein).

Quantum fluctuations create a virtual particle pair near the BH horizon .
While the particle with negative energy tunnels into the horizon
(absorption), the other one having positive energy flies off into spatial
infinity (emission) and produces the HR. Applying the WKB approximation, the
emission and absorption probabilities of the tunneling particles give the
tunneling rate $\Gamma$\ as \cite{Srinivasan,Kruglov,Kruglov1}

\begin{equation}
\Gamma=\frac{P_{emission}}{P_{absorption}}=\exp(-2ImS)=\exp\left( -%
\frac{E_{net}}{T}\right) ,   \label{2}
\end{equation}

where $S$ is the action of the classically forbidden trajectory of the
tunneling particle, which has a net energy $E_{net}$ and temperature $T$.
One of the methods for finding $S$ is to use the HJ method. This method is
generally performed by substituting a suitable ansatz, considering the
symmetries of the spacetime, into the relativistic HJ equation. The
resulting radial integral always possesses a pole located at the event
horizon. However, using the residue theory the associated pole can be
analytically evaded \cite{Feynman}.

Recently, within the framework of the HJ method, the HR of spin$-1$
particles described by the Proca equation in $3D$ non-rotating static BHs
has been studied by Kruglov \cite{Kruglov}. These spin$-1$ particles are in
fact the vector particles like $Z$ and $W^{\pm}$ bosons, and they have
significant role in the Standard Model \cite{SM}. Based on Kruglov's study 
\cite{Kruglov}, Chen et al. \cite{Chen} have very recently investigated the
HR\ of these bosons in the rotating BTZ geometry. Similar to the works of 
\cite{Kruglov,Chen}, here we aim to study the HR\ of the vector particles in
the $3D$ RBHWSH \cite{Zhao1,Zhao2,Zou,Belhaj}. These BHs are solutions to
the action in $3D$ Einstein gravity that is non-minimally coupled to scalar
field $\phi.$ In the limit of $\phi=0$, RBHWSH is nothing but the rotating
BTZ BH \cite{BTZ,Zhao1}.

This paper is organized as follows. Sec. 2 introduces the geometrical and
thermodynamical features of the $3D$ RBHWSH spacetime. In Sec. 3, we study
the Proca equation for a massive boson in this geometry. Then, we employ the
HJ method with the separation of variables technique to obtain the HR of the
RBHWSH. Finally, in Sec.4, we present our remarks.

\section{$3D$ RBHWSH Spacetime}

The action in a $3D$ Einstein gravity with a non-minimally coupled scalar
field reads \cite{Zhao1}

\begin{equation}
\mathcal{I}=\frac{1}{2}\int{d^{3}x\sqrt{-g}}\left[ {R-g^{\mu\nu}\nabla_{\mu
}\phi\nabla_{\nu}\phi-\xi R\phi^{2}-2V(\phi)}\right] ,   \label{3}
\end{equation}

where the coupling strength $\xi$ between gravity and the scalar field is $%
1/8$. Furthermore, the scalar potential $V(\phi)$\ is given by 
\begin{equation}
V(\phi)=-\Lambda+\frac{1}{512}\left( \Lambda+\frac{\beta}{B^{2}}\right)
\phi^{6}+\frac{1}{512}(\frac{a^{2}}{B^{4}})\frac{\left( \phi^{6}-40\phi
^{4}+640\phi^{2}-4608\right) \phi^{10}}{(\phi^{2}-8)^{5}},   \label{4n}
\end{equation}
in which the parameters $\beta$, $B$, and $a$ are integration constants, and 
$\Lambda$ is the cosmological constant. The line-element of RBHWSH is given
by 
\begin{equation}
\mathrm{d}s^{2}=-f(r)\mathrm{d}t^{2}+\frac{1}{f(r)}\mathrm{d}r^{2}+r^{2}%
\left[ \mathrm{d}\theta+\omega(r)\mathrm{d}t\right] ^{2},   \label{5}
\end{equation}
with the metric functions:

\begin{equation}
f(r)=-M\left( 1+\frac{2B}{3r}\right) +r^{2}\Lambda+\frac{\left( 3r+2B\right)
^{2}J^{2}}{36r^{4}},   \label{6n}
\end{equation}%
\begin{equation}
\omega(r)=-\frac{\left( 3r+2B\right) J}{6r^{3}},   \label{7n}
\end{equation}

where $J$ is the angular momentum of the BH. The scalar field is represented
by

\begin{equation}
\phi=\pm\sqrt{\frac{8B}{r+B}}.   \label{8n}
\end{equation}

It is worth noting that RBHWSH can be reduced to the rotating BTZ BH
solution when $B=0$ \cite{Zhao1,Zhao2,Zou,Belhaj}. \ Following \cite%
{Zou,Belhaj}, one can see that the mass, the Hawking\ temperature, the
Bekenstein-Hawking entropy, and the angular velocity of the particle at the
horizon of this BH are given by

\begin{equation}
M=\frac{J^{2}l^{2}\left( 2B+3r_{+}\right) ^{2}+36r_{+}^{6}}{%
12l^{2}r_{+}^{3}\left( 2B+3r_{+}\right) },   \label{9n}
\end{equation}

\begin{equation}
T_{H}=\frac{f^{\prime}(r_{+})}{4\pi}=\frac{(B+r_{+})\left[
36r_{+}^{6}-J^{2}l^{2}(2B+3r_{+})^{2}\right] }{24\pi l^{2}r_{+}^{5}\left(
2B+3r_{+}\right) },   \label{10n}
\end{equation}

\begin{equation}
S_{BH}=\frac{A_{H}}{4G}[1-\xi\phi^{2}(r_{+})]=\frac{4\pi r_{+}^{2}}{B+r_{+}}%
,   \label{11n}
\end{equation}

$\ $%
\begin{equation}
\Omega_{H}=-\omega(r_{+})=\frac{\left( 3r_{+}+2B\right) J}{6r_{+}^{3}}, 
\label{12n}
\end{equation}

where $\Lambda=\frac{1}{l^{2}},$ and $r_{+}$ is referred to as the event
horizon of the BH. In order to carry out analysis into finding $r_{+}$
values, we impose the condition of $f(r_{+})=0$, which yields a particular
cubic equation. The solutions of that cubic equation are also given in
detail by \cite{Zou}. One can verify that the first law of thermodynamics:

\begin{equation}
dM=T_{H}dS_{BH}+\Omega_{H}dJ.   \label{13n}
\end{equation}

holds. On the other hand, calculation of the specific heat using $%
C_{J}=T_{H}\left( \frac{\partial S_{BH}}{\partial T_{H}}\right) _{J}$ proves
that RBHWSH is locally stable when $r_{+}>r_{ext}$. Here, $r_{ext}$
represents the radius of an extremal RBHWSH that yields $T_{H}=0$ \cite{Zou}.

\section{HR of Spin-1 Particles From RBHWSH}

As described by Kruglov \cite{Kruglov}, the Proca equation for the massive
vector particles having the wave function $\Phi$ is given by

\begin{equation}
\frac{1}{\sqrt{-g}}\partial_{\mu}(\sqrt{-g}\Phi^{\nu\mu})+\frac{m^{2}}{%
\hbar^{2}}\Phi^{\nu}=0,   \label{14n}
\end{equation}

where

\begin{equation}
\Phi_{\nu\mu}=\partial_{\nu}\Phi_{\mu}-\partial_{\mu}\Phi_{\nu}. 
\label{15n}
\end{equation}

Let us set the vector function as 
\begin{equation}
\Phi_{\nu}=\left( c_{0},c_{1},c_{2}\right) \exp\left[ \frac{i}{\hbar }%
S(t,r,\theta)\right] ,   \label{16n}
\end{equation}
and assume that the action is given by 
\begin{equation}
S(t,r,\theta)=S_{0}(t,r,\theta)+\hbar
S_{1}(t,r,\theta)+\hbar^{2}S_{2}(t,r,\theta)+....   \label{17}
\end{equation}
According to the WKB\ approximation, we can further set 
\begin{equation}
S_{0}(t,r,\theta)=-Et+L(r)+j\theta+\Bbbk,   \label{18}
\end{equation}
where $E$\ and $j$ are the energy and angular momentum\ of the spin-1
particles, respectively, and $\Bbbk$ is a (complex) constant. Substituting
Eqs. (15), (16), (17), and (18) into Eq. (14) and considering the leading
order in $\hbar$, we obtain a $3\times3$ matrix (let us say $\Xi$ matrix)
equation:\ $\Xi\left( c_{1},c_{2},c_{3}\right) ^{T}=0$ (the superscript $T$
means the transition to the transposed vector). Thus, one can read the
non-zero components of $\Xi$ as follows

\begin{align}
\Xi_{11} & =A_{1}-j^{2},\   \notag \\
\Xi_{12} & =\Xi_{21}=-A_{2}\partial_{r}L(r),  \notag \\
\Xi_{13} & =\Xi_{31}=A_{1}\omega(r)-jE,\   \notag \\
\Xi_{22} & =\left( m^{2}r^{2}+j^{2}\right) f(r)-A_{2},  \notag \\
\Xi_{23} & =\Xi_{32}=\partial_{r}L(r)\left[ A_{2}%
\omega(r)^{2}-jf(r)^{2}\right] ,\   \notag \\
\Xi_{33} & =\frac{A_{1}}{r^{2}}\left[ f(r)-\omega(r)^{2}r^{2}\right] -E^{2}, 
\label{19}
\end{align}

where

\begin{equation}
A_{1}=r^{2}\left\{ m^{2}+f(r)\left[ \partial_{r}L(r)\right]
^{2}\right\} ,   \label{20n}
\end{equation}

\begin{equation}
A_{2}=r^{2}f(r)\left[ E+j\omega(r)\right] .   \label{21n}
\end{equation}

Upon the fact that any homogeneous system of linear equations (19) admits
nontrivial solution if and only if $det\Xi=0,$ we obtain

\begin{equation}
det\Xi=\frac{m^{2}}{r^{6}}[A_{1}+j^{2}-\frac{A_{2}^{2}}{r^{2}f(r)^{3}}%
]^{2}=0.   \label{22}
\end{equation}

Solving for $L(r)$ yields

\begin{equation}
L_{\pm}(r)=\pm\int\sqrt{\frac{\left[ E+\omega(r)j\right]
^{2}-f(r)(m^{2}+\frac{j^{2}}{r^{2}})}{f(r)^{2}}}dr.   \label{23}
\end{equation}
One can immediately observe from the above that when $\omega(r)=0$, it
reduces to the Kruglov's solution \cite{Kruglov}. The $L_{+}$
corresponds to outgoing (moving away from the BH) spin-1 particles and $%
L_{-}$ stands for the ingoing (moving towards the BH) spin-1\
particles. The imaginary part of $L_{\pm}(r)$ can be calculated by
using the pole deployed at the horizon. According to the complex path
integration method via the Feynman's prescription \cite{Feynman} ( see \cite%
{Srinivasan} for a similar process), we have

\begin{equation}
ImL_{\pm}(r)=\pm \frac{\pi}{f^{\prime}(r_{+})} E_{net}, 
\label{24n}
\end{equation}

where

\begin{equation}
E_{net}=E+E_{0}=E+\omega(r_{+})j=E-j\Omega_{H}.   \label{25n}
\end{equation}

Therefore, the probabilities of the vector particles crossing the horizon
out/in become

\begin{align}
P_{emission}& =\exp \left( -\frac{2}{\hbar }ImS\right) =\exp [-\frac{2%
}{\hbar }(ImL _{+}+Im\Bbbk )],  \label{26n} \\
P_{absorption}& =\exp \left( \frac{2}{\hbar }ImS\right) =\exp [-\frac{%
2}{\hbar }(ImL _{-}+Im\Bbbk )].  \label{27}
\end{align}

According to the classical definition of the BH, any outside particle
certainly falls into the BH. So, we must have $P_{absorption}=1$, which
results in $Im\Bbbk =-ImL _{-}.$ On the other hand, $%
L _{+}=-L _{-}$ so that the total probability of
radiating particles (as a consequence of QM) is

\begin{equation}
\Gamma=P_{emission}=\exp\left( -\frac{4}{\hbar}ImL_{+}%
\right) =\exp\left( -\frac{4\pi}{f^{\prime}(r_{+})}E_{net}\right) . 
\label{28}
\end{equation}

Thus, comparing Eq. (28) with Eq. (2) we can recover the correct Hawking
temperature\ (10) of RBHWSH: 
\begin{equation}
T\equiv T_{H}=\frac{f^{\prime}(r_{+})}{4\pi}=\frac{(B+r_{+})\left[
36r_{+}^{6}-J^{2}l^{2}(2B+3r_{+})^{2}\right] }{24\pi l^{2}r_{+}^{5}\left(
2B+3r_{+}\right) }.   \label{29}
\end{equation}

\section{Conclusion}

In this paper, we have used the Proca equation to compute the tunneling rate
of outgoing vector particles across the event horizon of axially symmetric
static rotating $3D$ RBHWSHs. For this purpose, we have ignored the
back-reaction effects and substituted the HJ ansatz into the associated
Proca equations. In the derivation of the tunneling rate within the
framework of the WKB approximation, the calculation of the imaginary part of
the action was the most important point. Using the complex path integration
technique, we have shown that the tunneling rate is given by Eq. (26). The
latter result allowed us to recover the standard Hawking temperature for
RBHWSH.

Finally, a study about the vector particles in the higher dimensional BHs
may reveal more information compared to the present case. This is going to
be our next problem in the near future.

\end{document}